# Barkhausen noise in the organic ferroelectric copolymer P(VDF:TrFE)

*Andrey Alekseevich Butkevich, Marcel Hecker, Toni Seiler, Martijn Kemerink*

Andrey Alekseevich Butkevich, Marcel Hecker, Toni Seiler, Martijn Kemerink

Institute for Molecular Systems Engineering and Advanced Materials, Im Neuenheimer Feld 225, Heidelberg, 69120, Germany.

Email: martijn.kemerink@uni-heidelberg.de

Polarization reversal within a ferroelectric material is commonly described as a progression of smaller switching events, giving rise to crackling or Barkhausen noise. While studies on Barkhausen noise, and particularly the associated event size distribution, allow for better understanding of switching processes in ferroelectrics, they were not yet conducted experimentally on organic ferroelectric materials. In this work, Barkhausen noise in the organic ferroelectric copolymer poly(vinylidene fluoride-co-trifluoroethylene) (P(VDF:TrFE)) is experimentally investigated under different electric fields, increasing at various rates. A weak dependence of the structure of the Barkhausen noise on both the magnitude and rise time of the applied electric field is observed, which manifests as a trend in the probability density function power-law exponents. Specifically, an increase in maximum electric field leads to an increase of the power-law exponent; increasing the rise time causes a parallel shift towards lower exponents. While these findings do not allow to conclusively confirm or refute universal self-organized critical behavior of the polarization reversal avalanches in P(VDF:TrFE), the exponents were found to seemingly converge to the 'universal' value of 1.5 for fast and strong driving, suggesting the system is close to this limit.

## 1. Introduction

For organic ferroelectrics, it has been established that bistable polarization and hence ferroelectric switching can be achieved by multiple mechanisms, like reorientation of permanent dipoles generating spontaneous polarization (conventional ferroelectrics),[1] donor-acceptor systems (charge-transfer complexes),[2–5] hydrogen-bonded compounds,[6–9] and many others.[10,11] However, insight into how one bistable state changes into another is almost entirely lacking. The present explanations have been mostly derived from theory and developed models.[12–14] While analytical (macroscopic), Monte Carlo (microscopic) and first-principle theory based (density functional theory) models have been applied to describe the ferroelectric switching processes, molecular dynamics have especially been successful in this task.[14–18]. In contrast, experimentally measuring and investigating the underlying mechanisms has been proven to be challenging for both inorganic and especially organic materials.[19]

One powerful method to study the process of ferroelectric switching and gain more insight into the cooperativity and the disorder being characteristic of extrinsic switching is the observation of crackling – or Barkhausen – noise, which appears in ferroelectric materials during ferroelectric polarization reversal. It has been investigated and was confirmed during switching processes, although virtually all studies were confined to inorganics and especially perovskites.[20–26] Typically, in such noise spectroscopy experiments, power-law distributions were established for multiple quantities related to the switching processes. While the recovered exponents varied between different materials, the overall similar behavior suggests universality in systems exhibiting Barkhausen noise.[21–23,25,26]

For organic ferroelectric materials, the – to our best knowledge – only experimental investigation into Barkhausen noise was performed by Mai and Kliem on thin films of the well-known organic ferroelectric copolymer poly(vinylidene fluoride-co-trifluoroethylene) (P(VDF:TrFE)).[27] The chemical structure of P(VDF:TrFE) is given in **Figure 1a**. In their work, the measured current peaks were generated without an applied electric field due to the thermally-induced reordering of dipoles while crossing the Curie-temperature. Although the Barkhausen pulses only corresponded to the switching of a very small number of dipoles (much less than 1% of total dipoles), clear deviations from a continuous switching process with different shapes were observed.[27] While this study confirmed Barkhausen noise in the copolymer PVDF, it did neither investigate the system through ferroelectric switching nor the power-law behavior. Furthermore, we previously performed a study on the prototypical organic molecular ferroelectric benzene-1,3,5-tricarboxamide (BTA).[**REF**] While we were able to numerically model the BTA system via kinetic Monte Carlo simulations, indeed finding Barkhausen noise and strong indications for self-organized criticality, we could not measure the Barkhausen noise experimentally as the number of simultaneously switching dipoles was, in actual agreement with the simulations, not enough to provide measurable currents given the resolution threshold of the used setup. Prior to coming to the specific goals of the present investigation, we will first give a brief overview of relevant previous works on Barkhausen noise and switching kinetics in P(VDF-TrFE).

Physical systems typically exhibit specific length scales, characterizing the events occurring within them. In critical systems, the events occur on all possible length scales. The corresponding critical behavior often gives rise to a power-law distribution of the event sizes, which, however, is not sufficient as the only indicator for critical behavior but should be present in case of one.[28,29] If a power-law occurs in a critical system, the value of the exponent may indicate universality. Barkhausen noise may be used as a probe of such critical behavior. It is the response of a system to the change of external conditions such as electric field, temperature, etc. with suddenly appearing and discrete events of varying sizes. In ferroelectric materials, in the vicinity of the coercive field the change of external electric field causes (groups of) dipoles to switch and hence the polarization to change.[25] The resulting events are discrete, can vary widely in size and indeed seem to typically follow a power-law distribution.[30]

In general, not all systems exhibit crackling behavior. Depending on their internal structure, and especially the structural disorder, some respond to external forces with numerous similarly sized and small events, others have typically only few large events as their response. The former is also called popping while the latter is known as snapping. Systems exhibiting crackling noise fall between these two cases based on the interactions within the system, hence representing a transitional behavior between popping and snapping.[30] As such, noise spectroscopy can be used to investigate the internal structure and cooperativity in ferroic materials, with the (relative) strength of disorder determining the spectral distribution of the events – intrinsic switching in an ideal, defect-free ferroelectric would give rise to a single snapping event spanning the entire volume, whereas dominant disorder would wash out any correlations, leading to popping noise.

The mean field plasticity (MFP) model was initially developed for materials under shear stress which display step-like stress-strain curves and exhibit material-independent power-law event size distributions.[31] Similar to how crackling noise can manifest as popping, crackling or snapping, these materials show varied behaviors depending on the coupling of their individual parts. The MFP model assumes that a slip is initiated once the local failure stress $\tau_f$ is exceeded and continues until the local stress falls below the sticking stress $\tau_s$. Due to material disorder, both the local failure and sticking stress can vary depending on the location within the material. Furthermore, the initial slip may induce additional slips in adjacent areas as it alters the local stress, hence triggering an avalanche propagating through the material as the external force $F$ gradually increases. This description yields an equation of motion whose solution predicts, in accordance with the domain wall propagation model, a power-law distribution of all sizes with an exponential cut-off.[32] The theory also describes that the distributions are (stress) integrated over the applied force upon collecting the data over extensive stress ranges.[33] This integration causes the power-law exponents to increase as the largest avalanches occur at the critical stress.[34]

It was shown that the MFP model can also be applied to the switching processes in ferroic materials, in which case the step-like stress-strain curve is to be understood as the step-like motion of a domain boundary, resulting from the competition between, typically, a driving force in the form of an increasing electric field and stopping force associated with pinning at structural defects.[31,35] For ferroic materials, this implies that data collected over the entire hysteresis loop aligns with the stress integrated exponents, whereas data taken solely around the coercive field corresponds to the non-stress integrated values.[34] The resulting critical power-law exponents describing the event size distribution $\tau$ and energy distribution $\epsilon$ are 2 and 5/3 for the former and 3/2 and 4/3 for the latter, respectively.[31]

A seemingly unrelated description of the switching kinetics of real, i.e. disordered, ferroelectrics is provided by the thermally activated nucleation limited switching (TA-NLS) model that states that switching processes in ferroelectric materials are limited and initiated by the activation of pre-existing small nucleation sites.[36,37] The activation of a nucleation site may be described as the thermally driven polarization reversal of a critical volume $V^*$, which is then followed by further domain growth. The latter can in principle be left unspecified, but is commonly described by the classical Kolmogorov–Avrami–Ishibashi (KAI) theory that accounts for (continuous) space filling domain growth.[38,39]

The TA-NLS model provides the following expression for the coercive field $E_c$ dependent on the temperature $T$ and the voltage sweeping frequency $\nu_{\text{exp}}$:

$$E_c = \frac{w_b}{P_s} - \frac{k_B T}{V^* P_s} \ln\left(\frac{\nu_0}{\nu_{\text{exp}} \ln 2}\right), \quad (1)$$

with the energy barrier between two metastable states $w_b$, the saturation polarization $P_s$, the Boltzmann factor $k_B$ ($k_B T$ as thermal energy of the system) and the attempt-to-flip frequency associated with the material specific phonon frequency $\nu_0$. A high sweeping frequency in a hysteresis loop measurement leads to a large coercive field since the system cannot equilibrate and fully relax in time. Furthermore, the model predicts that the coercivity of a ferroelectric system increases with the driving frequency since the switching itself becomes more intrinsic, giving rise to more larger and less smaller switching events, similar to the behavior of the creep regime. The coercive field decreases linearly with temperature and the logarithm of the inverse frequency, as shown in **Equation 1**. In another work, we simulated the Barkhausen noise in the organic molecular discotic BTA and concluded that the power-law exponents are anti-correlated to the coercivity of the system.[**REF**]

The TA-NLS model has been applied to the nucleation sites in P(VDF:TrFE) and the critical volume $V^*$ was estimated to be 4 nm$^3$ in a previous work.[39] As discussed in **Section S1** in the Supplementary Information (SI), this volume yields a number of dipoles which is many orders of magnitude below of what can be measured experimentally by noise spectroscopy. Still, the resulting avalanches might be detectable, which is the purpose of the current work. In particular, while the (Preisach) distribution of nucleation sites in P(VDF:TrFE) was previously shown to give insight in how these domains interact, this gives no information about the subsequent polarization reversal cascades. Therefore, it is not clear whether a direct or indirect relation exists between the Preisach and Barkhausen event size distributions.

In this work, we demonstrate experimental measurements of Barkhausen noise in the ferroelectric copolymer P(VDF:TrFE). The measurements are conducted for multiple applied voltages below, at and above the coercive field as well as for three different rise times. The power-law exponents are extracted from the obtained histograms, which are fitted with two different fitting methods. Depending on the rise time, the power-law exponents sit around or slightly below the universal values for critical behavior predicted by the MFP model. Increasing the applied electric field from below to above the coercive field leads to a weak increase of power-law exponents. Longer rise times and hence slower driving shift the power-law exponents to lower values, which agrees with the TA-NLS model. While the trends of the power-law exponents do not suggest universality, they do, for higher fields and faster driving, converge around the estimated value of 1.5, suggesting the system is close to critical; for lower fields and slower driving, the system appears to be in a regime with an abundance of larger 'snapping' events.

## 2. Samples, data acquisition and analysis procedure

Measurements were carried out on out-of-plane samples with a thin P(VDF:TrFE) film between two crossed gold electrodes. Both side and top views of a single device are shown in **Figure 1b and c**. A schematic of the measurement setup is provided in **Figure S2.1** and further experimental details are given in **Section 5**. Prior to noise measurements, the ferroelectric switching of the samples was checked; a typical result and its explanation are shown in **Figure S2.2** in the SI. Values found for the coercive field $E_c$ and remnant polarization $P_r$ were (dependent on frequency, temperature and applied voltage) approximately $20 - 25$ V µm$^{-1}$ and $60 - 65$ mC m$^{-2}$, respectively, in good agreement with previous results.[40,41]

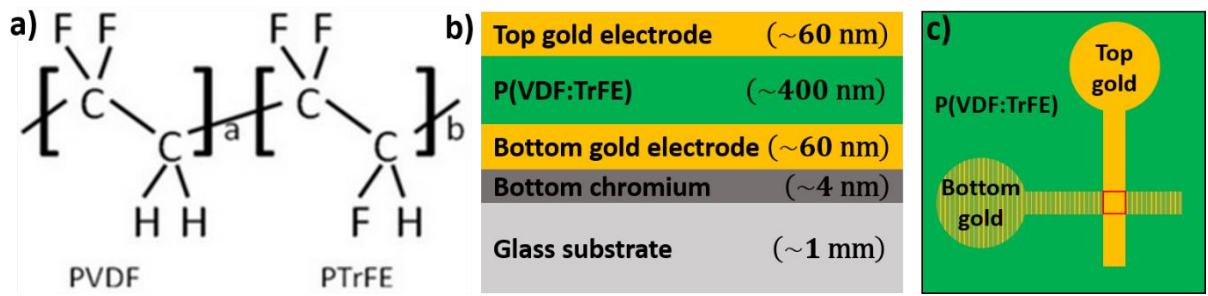

**Figure 1: a)** Chemical structure of the P(VDF:TrFE) random copolymer. The used ratio was 70% PVDF and 30% TrFE. **b)** Side and **c)** top views of an out-of-plane device used for ferroelectric characterization. The top view shows that the deposited electrodes have a well-defined cross-section (marked with a red square) where the electric field is applied. The bottom electrode is marked with stripes.

The measuring setup was optimized to provide the lowest possible noise floor with the given devices and is described in some detail in the **Section S3** in the SI. Every measurement, that is a single combination of applied voltage and rise time, was repeated multiple times for each sample and the data was captured in bundles of 100 current waveforms. The electric field was applied via the double wave method (DWM) which is depicted in Figure S2.2 and **Figure S2.3** in the SI. In short, the DWM applies a poling pulse twice in every poling direction, allowing to correct for background currents and hence isolate the switching current by subtracting the second signal peak from the first assuming that the first poling peak saturates the switching current for the given electric field strength. Since not all measurements exhibited Barkhausen noise, the data was cleaned up before analysis. For each measurement, the slew-rate $S(t) = \left(\frac{dI}{dt}\right)^2$ was calculated and compared to a threshold based on the maximum applied voltage. If the threshold was not met, the waveforms were rejected. Otherwise, the waveform was displayed and manually approved in order to further sort out spurious signals which appeared semi-randomly, but seemed to become more frequent after prolonged measurement, especially at higher fields. Although this is not pursued further here, we tentatively attribute this to a gradual decrease of structural disorder due to an effective 'field annealing'. Interestingly and importantly, the appearance or disappearance of the Barkhausen noise signals was not correlated to the overall switching behavior which continued to correspond to the full polarization reversal, happening around a constant coercive field. An automated approach was explored but was declined due to a high false-positive rate.

The sorted datasets were analyzed by multiple steps. An exemplary analysis process is shown in **Figure 2**. First, the calculated slew-rate $S(t) = \left(\frac{dI}{dt}\right)^2$ was considered and datasets with $S < 10^{-5} \frac{A^2}{s^2}$ were excluded. This threshold was based on typical measured results so that most of the background noise was excluded, as shown in **Figure 2b**. Next, the baseline was calculated by identifying the local minima of the slew-rate and interpolating these datapoints using the piecewise cubic Hermite interpolating polynomial (PCHIP). The interpolated baseline was subtracted from the calculated slew-rate to account for the underlying slew-rate caused by the slope of the switching peak, as indicated in **Figure 2c**. Using the corrected slew-rate, the probability density function (PDF) was calculated. Finally, power-laws were fitted on to the PDFs in double logarithmic representation of the data to obtain the power-law exponents via the slope of the fit, as shown in **Figure 2d**. For better statistics, two methods were used for the power-law fitting: The maximum likelihood (ML) and the least squares (LS) methods. Details on these fitting methods can be found in **Section S2** in the SI.

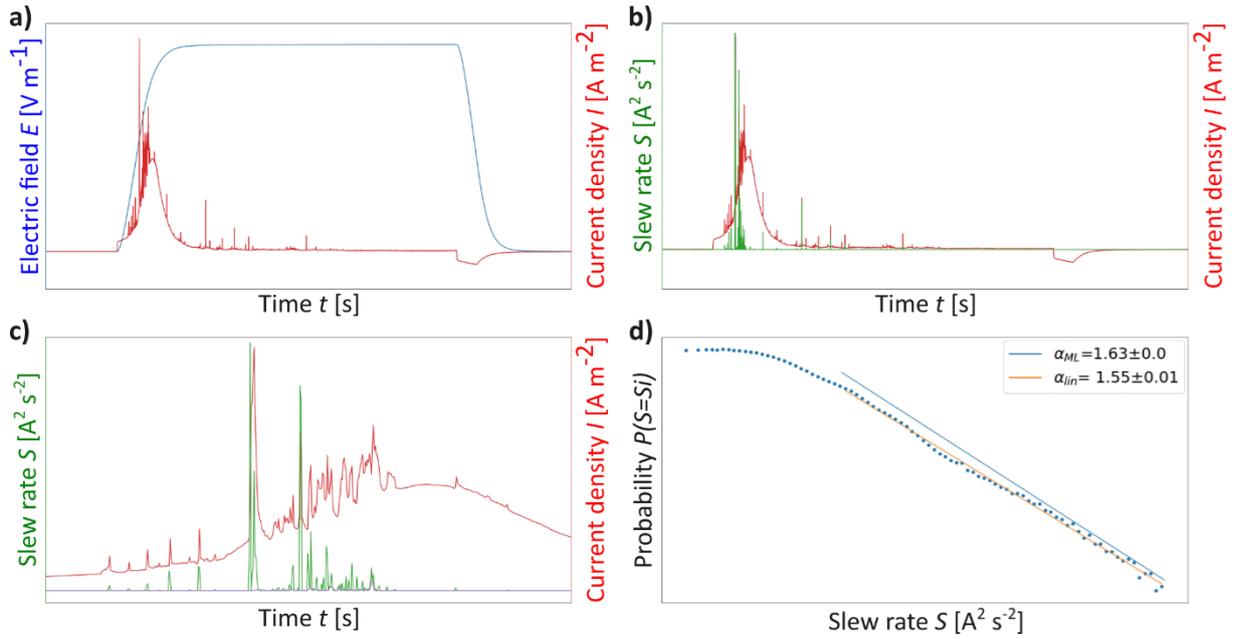

**Figure 2:** An example of a typical data analysis. The absolute values on the axes are left out here. **a)** The measured current density of a P(VDF:TrFE) sample (red) as a response to an applied square electric field waveform (blue). **b)** The slew rate $S$ (green) calculated from the current signal (red). **c)** The slew rate $S$ (green) and current signal (red) zoomed in on the relevant part of the measured spectrum with the interpolated slew rate baseline (purple). **d)** The fitted power-laws with the extracted power-law exponents for the measurement in (a) using both ML (blue line) and LS (orange line; depicted as linear) fitting methods.

### 3. Experimental results

Based on previous results, an effect of both the rise time and the applied voltage on the power-law exponents was expected. Thus, the measurements were done for three different rise times, cf. Section S3 in the SI, while maintaining a constant frequency of the total double-wave pulse train. Likewise, the applied voltage was varied from below to well beyond the coercive field.

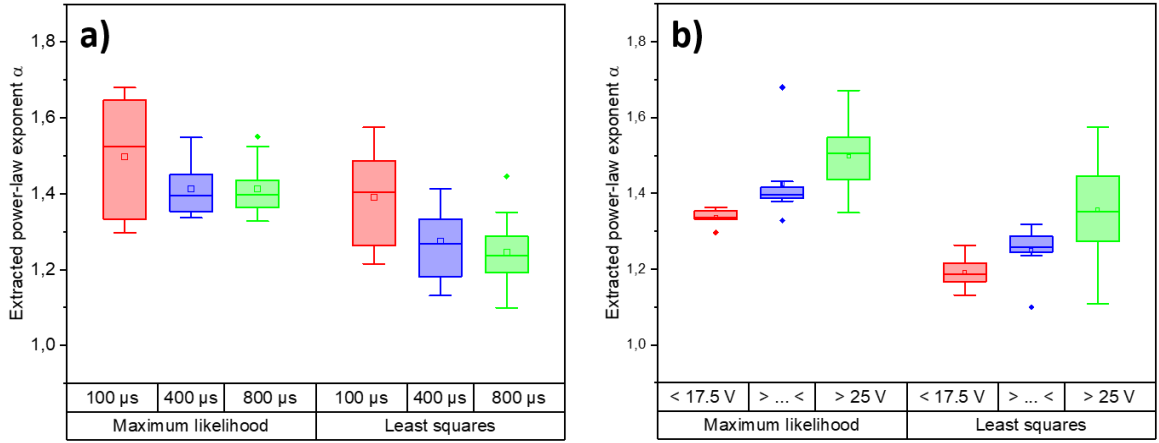

**Figure 3:** Box plot of the extracted critical power-law exponents $\alpha$ for **a)** three different rise times including all applied voltages for each rise time in the analysis and **b)** three different intervals of the applied voltage including all rise times for each voltage interval in the analysis for both fitting methods. The shorthand label of the middle panel indicates voltages between 17.5 and 25 V. Some data points were deemed inconclusive based on the difficulty to obtain unique fits to the histograms and were excluded here. A similar box plot with all values is shown in **Figure S4.5** in the SI.

The box plot in **Figure 3** summarizes our findings; the underlying histograms for all individual measurements are provided in **Section S4** (**Figure S4.2-4**) in the SI with a measurement of a dielectric reference capacitor provided in **Figure S4.1** for comparison. The variability of the samples during prolonged measurements that was discussed above also led to a significant fraction of non-ideal probability distributions, which inspired the use of two independent approaches, ML and LS, to determine the power law exponents. Apart from a small offset, the two approaches lead to similar values of the mean exponent, slightly below the mean field value of 1.5. In addition, as visible in **Figure 3a**, both analysis methods suggest a weak trend of increasing exponents for higher driving speeds (and hence lower rise times), with the 100 µs rise time leading to values at or close to the 1.5 expected from the mean field plasticity model. The same trend is visible for both fitting methods when plotting the power-law exponents against the applied voltage for the three different rise times, see **Figure 3b**. The individual exponent values for different applied voltages and rise times are shown in **Figure S4.6**. In addition, higher applied voltages, and hence higher electric fields, lead, for a fixed rise time, to higher power-law exponents with an increase of roughly $0.01\ V^{-1}$, as indicated by the dashed lines in **Figure S4.6**.

The trends in extracted power-law exponents both as a function of rise times and applied voltage suggest that the system is not truly critically self-organized, as no universal behavior seems to be present. However, the power-law exponent values for both fastest (100 µs) and strongest (above coercive field) driving are very close to the mean field value of 1.5, hence indicating that the system is close to such behavior. Furthermore, the distributions in general become more ideal for shorter rise times and higher fields (see Figure S4.2-4 in the SI), which is consistent with the notion that the system approaches self-organized criticality for those conditions. In contrast, the significant variability of the response curves and hence the extracted critical power-law exponents, especially at weaker driving, suggests that the behavior is not universally self-organized and instead depends on minor variations in the structural (dis)order that – for instance – occur upon repeated switching at higher fields. This, in turn, is in line with our observation of an increasing occurrence of spurious signals in prolonged measurements at higher fields, as mentioned previously in Section 2.

Furthermore, the trends in power-law exponents observed in Figure 3, Figure S4.5 and S4.6 are consistent with the TA-NLS model that predicts that increasing rise times, and hence lowering the driving speed, leads to less coercivity in the system, which yields fewer larger and more smaller events, resulting in generally smaller extracted power-law exponents. This is in line with our previous findings for coercivity and the corresponding event sizes within a simulated BTA ferroelectric system.[REF] Especially below the coercive field, the extracted power-law exponents are lower than expected from the mean field theory, which agrees with other studies carried-out for magnetic systems and can be explained by the entire system not being fully switched.[42] In that scenario, less stable parts of the system can switch before reaching the coercive field, leading to isolated smaller events that do not progress into larger, system-spanning avalanches and hence to smaller power-law exponents.

As mentioned in Section 1, the TA-NLS theory has previously been used to investigate the (Preisach) distribution of nucleation sites (hysterons) in P(VDF:TrFE), for which a typical critical volume $V^*$ of 4 nm$^3$, which corresponds to roughly $10^3$ switched dipoles, has been established.[39] In our case, the smallest event sizes which can be resolved experimentally correspond to roughly $10^8$ dipoles switching simultaneously. As such, the measured signals must reflect the resulting avalanches and not the nucleation events. Although the absence of specific data does not allow us to draw firm conclusions, we would like to propose a possible connection between the distributions of nucleation sites and switching events, inspired by our numerical work on BTA.[REF] In that work, we found that small hysterons, consisting of stacks of ~10 molecules each, that are separated from each other by morphological defects, can both act as nuclei for polarization reversal and form the building blocks for system-spanning avalanches. Whether this leads to universal critical behavior depends on the degree of, and variation in, the coupling between these hysterons. In the case of BTA, which is a quasi-1D system consisting of hexagonally packed supramolecular stacks, the maximum event size is nevertheless limited to several hundreds of molecules, that is, the size of a single column. Given its rather different morphology, domain growth in P(VDF:TrFE) may be expected to show a higher dimensionality that supports much larger avalanches that are still built up from much smaller, interacting hysterons,[38,39] which is supported by previous findings where the characteristic geometric dimension for the domain growth in P(VDF:TrFE) becomes generally larger on increasing the driving electric field.[43] While speculative, this may be the underlying mechanism for the herein observed behavior. The avalanches triggered by the small nuclei can, apparently, give rise to critical behavior that is close to self-organized critical. The degree to which that happens depends on how strongly the system is driven and, in view of the significant sample-to-sample variation and the absence of signals after longer operation, seems to depend on relatively subtle variations in morphology.

## 4. Summary

In this work, the Barkhausen noise associated with polarization reversal in thin film P(VDF:TrFE) samples was experimentally measured for different applied voltages and rise times. The obtained histograms of event sizes were fitted with both the maximum likelihood and least squares fitting methods and the power-law exponents were extracted. The exponents were consistently slightly higher for the former fitting method and below or at the mean field value of 1.5 for both fitting methods. Both decreasing the rise time and increasing the applied voltage led to an increase of the extracted power-law exponents. The trends of the extracted power-law exponents have been identified as consistent with the thermally-assisted nucleation-limited switching (TA-NLS) model. Based on the trends, it has been concluded that the system is not truly critically self-organized due to the lack of universal behavior. However, as the exponents for both the fastest rise time and applied voltages above the coercive field were close to the mean field value and the corresponding distributions became more ideal, it seems that the system approached self-organized criticality for those conditions. For weaker and slower driving, the behavior is not universally self-organized and instead depends on variations in structural (dis)order that varies between samples and even during measurements. Lastly, on basis of their magnitude, the measured Barkhausen noise cannot directly be mapped upon the distribution of nucleation sites in the TA-NLS model was deemed impossible. Instead, we speculate that the domain growth in P(VDF:TrFE) progresses through avalanches that are built up from larger numbers of such sites, with an interaction that strongly depends on sample morphology and (switching) history.

## 5. Experimental details

**Measuring setup:** The input signal was supplied by a Keysight 33600A Arbitrary Waveform Generator (AFG) and amplified by a TReK PZD350A high voltage amplifier. Considerations and tests that were performed to select these devices are provided in Section S3 in the SI. The device response was measured by either a Zurich Instruments impedance analyzer (MFIA) or lock-in amplifier (MFLI). No significant differences between the MFIA and MFLI were established. The device under test was measured inside a Linkam miniature cryostat featuring electric contacts to reduce noise and allow for precise temperature dependent measurements.

**Sample preparation:** Out-of-plane electrodes were used for device fabrication of thin film P(VDF:TrFE) samples. Commercially available Corning plain microscope slides with a thickness of 0.96 to 1.06 mm were cut and both bottom and top electrodes were thermally evaporated. Gold (on chromium in case of bottom) electrodes of different thicknesses (typically $50 - 70$ nm) were utilized. The P(VDF:TrFE) thin films (70:30 PVDF:TrFE) were spin coated with 900 rpm, yielding typical film thicknesses of ~400 nm which were determined via a stylus profilometer (Bruker DektakXT). Each sample consisted of a total of nine devices. Three different electrode widths were used for both top and bottom contacts (0.2, 0.5 and 1 mm), so that each sample included all possible nine combinations of bottom and top electrode widths. The results for the different electrode combinations were consistent, although combinations of larger electrode widths were significantly more prone to short-circuits. A total of 144 crossbar out-of-plane samples were tested, of which most were short-circuited and 32 showed Barkhausen noise.

## Acknowledgements

We thank Fabian Thome for contributions to the early stages of this work. We thank the Deutsche Forschungsgemeinschaft (DFG, German Research Foundation) for support of this work (SFB 1249 and EXC-2082/1-390761711). M.K. thanks the Carl Zeiss Foundation for financial support.

# References

[1] C. F. C. Fitié, W. S. C. Roelofs, P. C. M. M. Magusin, M. Wübbenhorst, M. Kemerink, R. P. Sijbesma, *J. Phys. Chem. B* **2012**, *116*, 3928.
[2] S. Horiuchi, T. Hasegawa, Y. Tokura, *J. Phys. Soc. Jpn.* **2006**, *75*, 051016.
[3] S. Horiuchi, K. Kobayashi, R. Kumai, S. Ishibashi, *Chemistry Letters* **2014**, *43*, 26.
[4] S. Horiuchi, S. Ishibashi, Y. Tokura, in *Organic Ferroelectric Materials and Applications* (Ed.: K. Asadi), Woodhead Publishing, **2022**, pp. 7–46.
[5] S. Barman, A. Pal, A. Mukherjee, S. Paul, A. Datta, S. Ghosh, *Chemistry – A European Journal* **2024**, *30*, e202303120.
[6] Z. Sun, T. Chen, C. Ji, S. Zhang, S. Zhao, M. Hong, J. Luo, *Chem. Mater.* **2015**, *27*, 4493.
[7] S. Horiuchi, S. Ishibashi, *J. Phys. Soc. Jpn.* **2020**, *89*, 051009.
[8] J. Wu, Q. Zhu, T. Takeda, N. Hoshino, T. Akutagawa, *ACS Appl. Electron. Mater.* **2021**, *3*, 3521.
[9] S. Horiuchi, S. Ishibashi, Y. Tokura, in *Organic Ferroelectric Materials and Applications* (Ed.: K. Asadi), Woodhead Publishing, **2022**, pp. 47–84.
[10] S. Horiuchi, Y. Tokura, *Nature Mater* **2008**, *7*, 357.
[11] H. Liu, Y. Ye, X. Zhang, T. Yang, W. Wen, S. Jiang, *Journal of Materials Chemistry C* **2022**, *10*, 13676.
[12] T. D. Cornelissen, M. Biler, I. Urbanaviciute, P. Norman, M. Linares, M. Kemerink, *Phys. Chem. Chem. Phys.* **2019**, *21*, 1375.
[13] T. D. Cornelissen, I. Urbanaviciute, M. Kemerink, *Phys. Rev. B* **2020**, *101*, 214301.
[14] T. Cornelissen, M. Kemerink, in *Organic Ferroelectric Materials and Applications* (Ed.: K. Asadi), Woodhead Publishing, **2022**, pp. 185–232.
[15] J. Liu, W. Chen, B. Wang, Y. Zheng, *Materials* **2014**, *7*, 6502.
[16] M. Sepliarsky, A. Asthagiri, S. R. Phillpot, M. G. Stachiotti, R. L. Migoni, *Current Opinion in Solid State and Materials Science* **2005**, *9*, 107.
[17] R. E. Cohen, in *Ferroelectricity*, John Wiley & Sons, Ltd, **2005**, pp. 139–154.
[18] U. V. Waghmare, *Acc. Chem. Res.* **2014**, *47*, 3242.
[19] A. Picinin, M. H. Lente, J. A. Eiras, J. P. Rino, *Phys. Rev. B* **2004**, *69*, 064117.
[20] V. Ya. Shur, E. L. Rumyantsev, D. V. Pelegov, V. L. Kozhevnikov, E. V. Nikolaeva, E. L. Shishkin, A. P. Chernykh, R. K. Ivanov, *Ferroelectrics* **2002**, *267*, 347.
[21] I. S. Baturin, M. V. Konev, A. R. Akhmatkhanov, A. I. Lobov, V. Ya. Shur, *Ferroelectrics* **2008**, *374*, 136.
[22] V. Ya. Shur, A. R. Akhmatkhanov, I. S. Baturin, E. V. Shishkina, *Journal of Applied Physics* **2012**, *111*, 014101.
[23] C. D. Tan, C. Flannigan, J. Gardner, F. D. Morrison, E. K. H. Salje, J. F. Scott, *Phys. Rev. Mater.* **2019**, *3*, 034402.
[24] C. Flannigan, C. D. Tan, J. F. Scott, *J. Phys.: Condens. Matter* **2019**, *32*, 055403.
[25] E. K. H. Salje, D. Xue, X. Ding, K. A. Dahmen, J. F. Scott, *Phys. Rev. Mater.* **2019**, *3*, 014415.
[26] B. Casals, G. F. Nataf, E. K. H. Salje, *Nat Commun* **2021**, *12*, 345.
[27] M. Mai, H. Kliem, *Journal of Applied Physics* **2013**, *114*, 224104.
[28] K. G. Wilson, *Scientific American* **1979**, *241*, 158.
[29] D. Sornette, *Critical Phenomena in Natural Sciences: Chaos, Fractals, Selforganization and Disorder: Concepts and Tools*, Springer Science & Business Media, **2006**.
[30] J. P. Sethna, K. A. Dahmen, C. R. Myers, *Nature* **2001**, *410*, 242.
[31] K. A. Dahmen, Y. Ben-Zion, J. T. Uhl, *Phys. Rev. Lett.* **2009**, *102*, 175501.
[32] K. A. Dahmen, in *Avalanches in Functional Materials and Geophysics* (Eds.: E. K. H. Salje, A. Saxena, A. Planes), Springer International Publishing, Cham, **2017**, pp. 19–30.

[33] M. Zaiser, B. Marmo, P. Moretti, in *Proceedings of International Conference on Statistical Mechanics of Plasticity and Related Instabilities — PoS(SMPRI2005)*, Sissa Medialab, Indian Institute of Science, Bangalore, India, **2006**, p. 053.

[34] G. Durin, S. Zapperi, *J. Stat. Mech.* **2006**, *2006*, P01002.

[35] E. K. H. Salje, K. A. Dahmen, *Annu. Rev. Condens. Matter Phys.* **2014**, *5*, 233.

[36] M. Vopsaroiu, J. Blackburn, M. G. Cain, P. M. Weaver, *Phys. Rev. B* **2010**, *82*, 024109.

[37] M. Vopsaroiu, P. M. Weaver, M. G. Cain, M. J. Reece, K. B. Chong, *IEEE Transactions on Ultrasonics, Ferroelectrics, and Frequency Control* **2011**, *58*, 1867.

[38] A. V. Gorbunov, T. Putzeys, I. Urbanavičiūtė, R. A. J. Janssen, M. Wübbenhorst, R. P. Sijbesma, M. Kemerink, *Phys. Chem. Chem. Phys.* **2016**, *18*, 23663.

[39] I. Urbanavičiūtė, T. D. Cornelissen, X. Meng, R. P. Sijbesma, M. Kemerink, *Nat. Commun.* **2018**, *9*, 4409.

[40] N. Meng, R. Mao, W. Tu, X. Zhu, R. M. Wilson, E. Bilotti, M. J. Reece, *Polymer* **2016**, *100*, 69.

[41] I. Urbanaviciute, *Multifunctional Supramolecular Organic Ferroelectrics*, Linköping University Electronic Press, **2019**.

[42] J. P. Sethna, K. A. Dahmen, O. Perkovic, **2005**, DOI 10.48550/arXiv.cond-mat/0406320.

[43] W. J. Hu, D.-M. Juo, L. You, J. Wang, Y.-C. Chen, Y.-H. Chu, T. Wu, *Sci Rep* **2014**, *4*, 4772.

Supporting information to

**Barkhausen noise in the organic ferroelectric copolymer P(VDF:TrFE)**

*Andrey Alekseevich Butkevich, Marcel Hecker, Toni Seiler, Martijn Kemerink*

Andrey Alekseevich Butkevich, Marcel Hecker, Toni Seiler, Martijn Kemerink

Institute for Molecular Systems Engineering and Advanced Materials, Im Neuenheimer Feld 225, Heidelberg, 69120, Germany.

Email: martijn.kemerink@uni-heidelberg.de

**Contents**

## S1. Estimation of the number of switched dipoles in P(VDF:TrFE)

From sample geometry, the total number of switchable dipoles within a material $N_\text{t}$ can be expressed as

$$N_\text{t} = \frac{P_\text{s} A_\text{s} D}{qd}. \tag{S1.1}$$

Assuming a saturation polarization of $P_\text{s} = 110$ mC m$^{-2}$,[1] a typical out-of-plane device with an active area of $A_\text{s} = 0.25$ mm $\times$ 0.25 mm, a dipole moment of $D = 1.8$ Debye,[2] the elementary charge $q$ and the size of a macroscopic dipole $d$, **Equation S1.1** leads to a total of $N_\text{t} \approx 4.6 \cdot 10^{14}$ dipoles contributing to the switching of a typical out-of-plane P(VDF:TrFE) device. The minimal resolvable peak for P(VDF-TrFE) assuming a minimal detectable current of 4 nA for the optimal required sampling time of 1 µs (938 kHz sampling rate) results in $\sim 2.8 \cdot 10^8$ dipoles as the minimal number of simultaneously switching dipoles required for detection. Meanwhile, a critical volume of 4 nm$^3$ obtained from TA-NLS theory yields a magnitude of $\sim 10^3$ dipoles, not allowing a direct experimental observation in noise spectroscopy.

## S2. Details on data acquisition and analysis

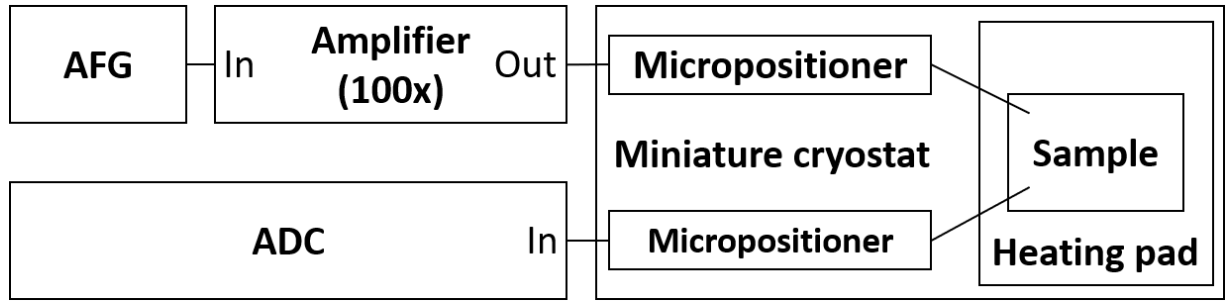

**Figure S2.1:** A schematic of the used measurement setup. The AFG is steered by a PC and applies a varying waveform signal which is amplified by an amplifier in order to reach the needed electric field strengths to the sample via contacting needle. The sample lies within a miniature cryostat which also allows to cool down and heat up the sample and simultaneously acts as a Faraday cage. The response voltage/current is measured by an analog-to-digital converter (ADC) which allows to trigger the signal and transfers it back to the PC for further analysis.

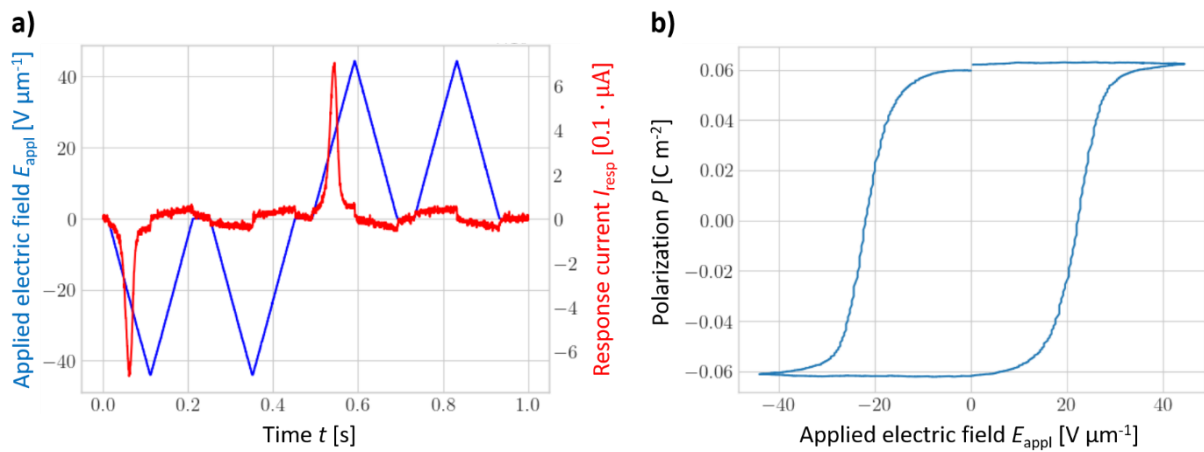

**Figure S2.2:** A typical switching current and the corresponding P-E-hysteresis loop of an out-of-plane P(VDF:TrFE) sample measured with a triangular double wave signal at 1 Hz and room temperature. **a)** The applied double wave voltage signal in both down and up directions (blue) and the corresponding sample response current (red). Here, the double wave is of triangular shape in order to obtain the hysteresis loop, as a truly square waveform does not allow to probe the response between the initial and final field values. In general, double wave signals allow to differentiate between the desired switching current and other contributions (like displacement and leakage) to the total current by subtracting the second (up or down) peak from the first.[3,4] Although this is necessary to obtain hysteresis loops as lossy dielectrics can exhibit hysteresis-like behavior[5], it is not needed for Barkhausen noise measurements as only the switching peak

appearing at the first waveform peak is considered, so that a single waveform (as in PUND – positive up negative down) could be used instead. **b)** The corresponding polarization hysteresis loop obtained from the background corrected and integrated switching peaks.

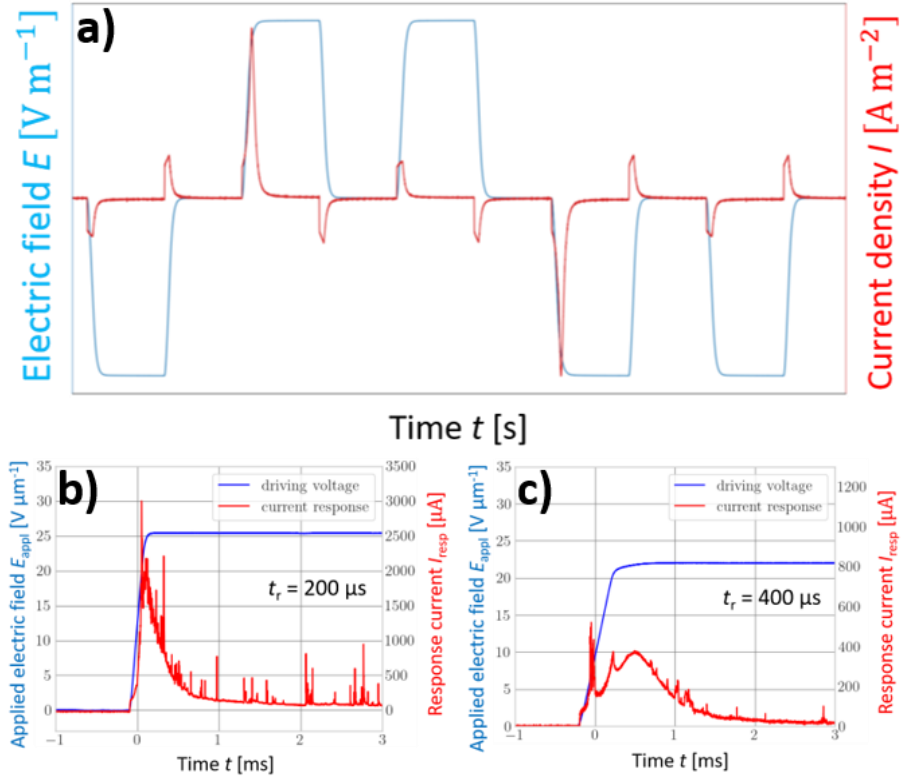

**Figure S2.3: a)** A measurement of current density of a P(VDF:TrFE) sample (red) in response to the applied electric field in form of a squared double wave (blue). Here, more than one measurement period is shown, the response current does not show Barkhausen noise and the absolute values on the axes are left out. The applied wave form is a square double wave with variable rise times, meaning that the step-like increase is not instantaneous. The rounding of the applied electric field arises due to the used (input) noise-filtering circuit and has no effect on the measured response current from the sample. Same measurements with Barkhausen noise (sudden increases of the current response) showcasing the changes of the shape of the main switching peak depending on the ramp time of the square waveform for 200 and 400 µs rising times are shown in **(b)** and **(c)**, respectively.

**Maximum likelihood fitting method:**

For a power-law starting at a value $x_{\min}$ and having an exponent $\alpha$, the probability $p(x)$ can be obtained via

$$p(x) = \frac{\alpha - 1}{x_{\min}} \left(\frac{x}{x_{\min}}\right)^{-\alpha}. \tag{S2.1}$$

If there are $N$ observations with $x_i \geq x_{\min}$, the probability of the data fitting the model is proportional to

$$p(x|\alpha) = \prod_{i=1}^{N} \frac{\alpha - 1}{x_{\min}} \left(\frac{x}{x_{\min}}\right)^{-\alpha}. \tag{S2.2}$$

The probability $p(x|\alpha)$ is the so-called likelihood. Maximizing the likelihood yields a better fit. Instead of the likelihood, the log-likelihood $\mathcal{L}$ – the logarithm of the likelihood – is commonly used. Considering the condition $\frac{\partial \mathcal{L}}{\partial \alpha} = 0$, the maximum likelihood estimator $\hat{\alpha}$ is obtained [6]:

$$\hat{\alpha} = 1 + N \left[\sum_{i=1}^{N} \ln\left(\frac{x}{x_{\min}}\right)\right]^{-1}. \tag{S2.3}$$

The corresponding calculations were implemented into a python package by Alstott et al. in 2014 and were used here for analysis and evaluation.[7]

**Least squares (linear) method:**

The aim of this probably most commonly used method for fitting models to data is to minimize the following expression:

$$\sum_{i=1}^{N} (y_i - f(x_i))^2. \tag{S2.4}$$

While the minimization can be done analytically for simple functions, the approach is typically non-trivial for non-linear functions. Many different algorithms exist to solve this problem. Within of this publication, the Levenberg-Marquardt algorithm was used.[8,9] A linear function $f(x) = mx + b$ is fitted to the measured data in double-logarithmic representation. Hence, the exponential of the fit function is plotted, which results in $f(x) = x^m e^b$, so that $m$ is associated with the exponent $\alpha$ and $e^b$ is just a constant.

## S3. Experimental details

**Amplifier characterization**

Two amplifiers (TReK PZD350A and Falco System WMA-200) were tested, of which the latter was found to have the lowest noise level. To check that the provided amplification is the constant within the used frequency range, the specified amplification and bandwidth were verified by measuring the amplification at various frequencies using a sine wave input generated by an arbitrary function generator (AFG). Measurements both with and without a capacitor (1 µF) were done and the corresponding schematic as well as measurement results are shown in **Figure S3.1** and **Figure S3.2**, respectively.

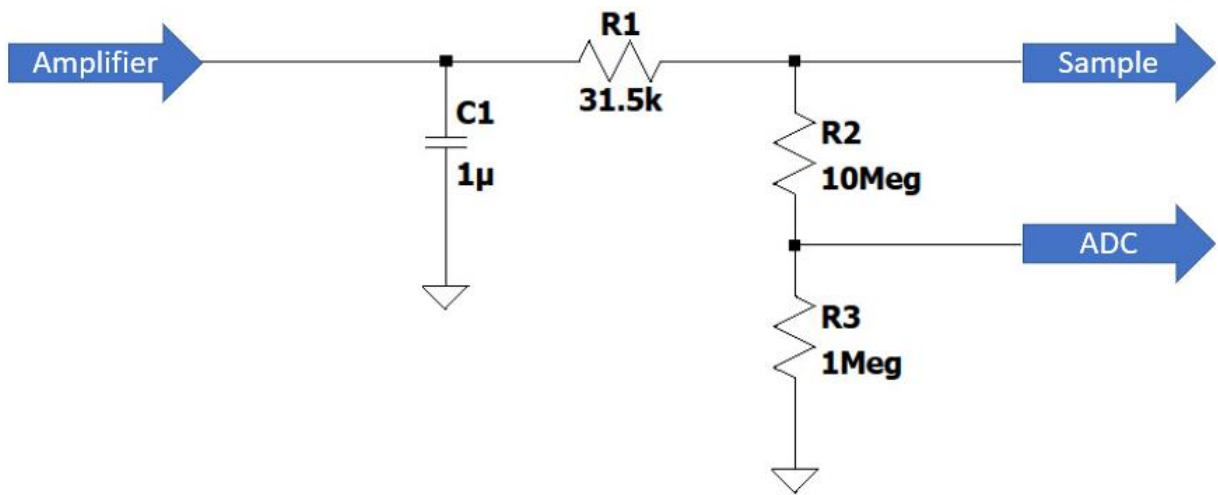

**Figure S3.1:** Circuit diagram of the measurement setup used for the amplifier characterization. Between the amplifier and the ADC, a RC-box is used in order to protect the instrument from excess voltage and current, which becomes especially relevant in case a sample short-circuits. It consists of three resistors of which one limits the current (R1) and the other two functioning as a voltage divider (R2 and R3). Additionally, a capacitor (C1) can be added to provide a base load to the amplifier, further reducing the noise level of the setup. The resistance values were chosen to obtain an advantageous division ratio considering the main voltage measurement input of 10 MΩ impedance and an auxiliary input of 1 MΩ impedance of the used ADCs.

The measured transfer function is fitted by the function

$$|H(f)| = \frac{G_o}{\sqrt{1 + \left(\frac{f}{f_c}\right)^2}} = \frac{|V_{in}|}{|V_{out}|} \quad (S3.1)$$

with the maximum gain $G_0$ and critical frequency $f_c$. The function in **Equation S3.1** is the magnitude of the transfer function $H(f)$ for first-order low-pass filters.[10] Figure S3.2 shows that the amplifier exceeds its specification for the frequency range (DC to 0.5 MHz) and is indeed sufficient for the conducted measurements. The slight drop in amplification at lower frequencies is caused by a built-in resistor which is in series with the measurement device, see Figure S3.1.

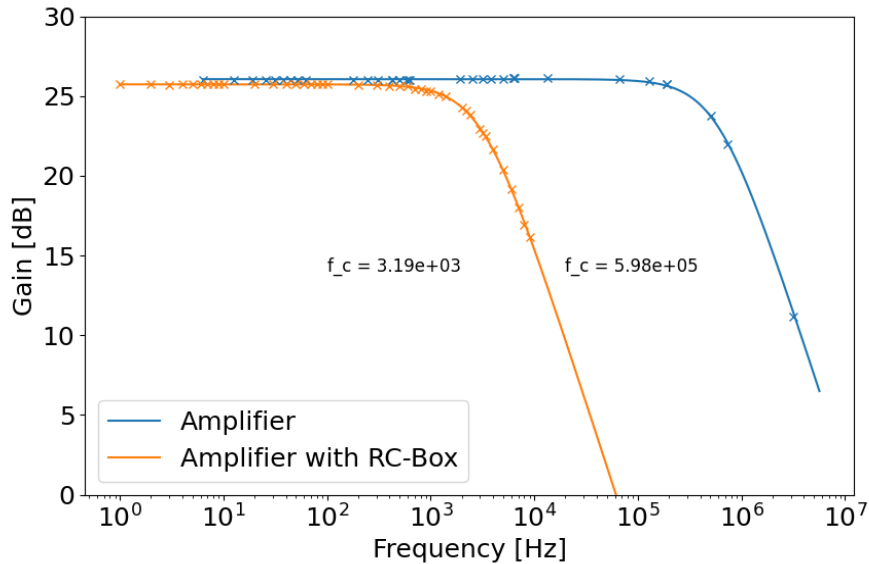

**Figure S3.2:** Measured amplification by TReK PZD350A high voltage amplifier of a sine signal generated by an AFG as a function of signal frequency $f$. The gain is plotted in decibels which equals to $20 \cdot \log_{10}(|H(f)|)$. The measurements (crosses) were conducted with (orange) and without (blue) a capacitor (1 µF) and fitted (lines) with the function depicted in Equation S3.1.

**Arbitrary function generator (AFG) characterization**

Multiple AFGs (Keysight 33600A, Tektronix AFG3052C and Tektronix AFG1062) were compared to determine which provides the most suitable performance for the conducted measurements. The measurements were done by using a reference setup: The AFG output was connected to a TReK PZD350A 20× amplifier which led to the sample which was substituted with a dielectric reference capacitor (10 pF, the typical sample capacitance is 6 to 35 pF) and finally into the input of a Zurich Instruments MFLI lock-in amplifier that is used as a high-performance analog-to-digital converter (ADC). The current passing through the capacitor is measured as the 1 MΩ input impedance of the MFLI is used as a current-to-voltage converter.

Since capacitors exhibit a current proportional to the displacement current, only current flow during constant voltage is considered. The MFIA's current range was set to 100 µA with a sampling rate of 937.5 kHz. The obtained noise numbers and measured noise spectra are shown in **Table S3.1** and **Figure S3.3**, respectively. The Keysight 33600A AFG was deemed to be the best suited for measuring Barkhausen noise. The measurement of Keysight 33600A AFG was repeated for multiple setup variations to confirm its best performance which is shown in **Figure S3.4**.

| AFG | Noise number [nA] |
|---|---|
| None (= noise floor of the MFIA/MFLI) | 1.65 ± 0.03 |
| Keysight 33600A | 2.57 ± 0.13 |
| Tektronix AFG3052C | 4.92 ± 0.03 |
| Tektronix AFG1061 | 12.41 ± 0.14 |

**Table S6.1:** Obtained noise numbers representing the noise levels of the tested AFGs. The measurements were done by using a reference setup in which the AFG output was connected to a 20× amplifier which led to the sample which was substituted by a dielectric reference capacitor with 10 pF (typical sample capacitance is 6 to 35 pF) and finally into the input of the Zürich Instruments lock-in (MFLI). Thus, the current passing through the capacitor is measured. Since capacitors exhibit a current proportional to the displacement current, only current flow during constant voltage is considered. The MFLI's current range was set to 100 µA with a sampling rate of 937.5 kHz. The noise number is calculated using the standard deviation of the current.

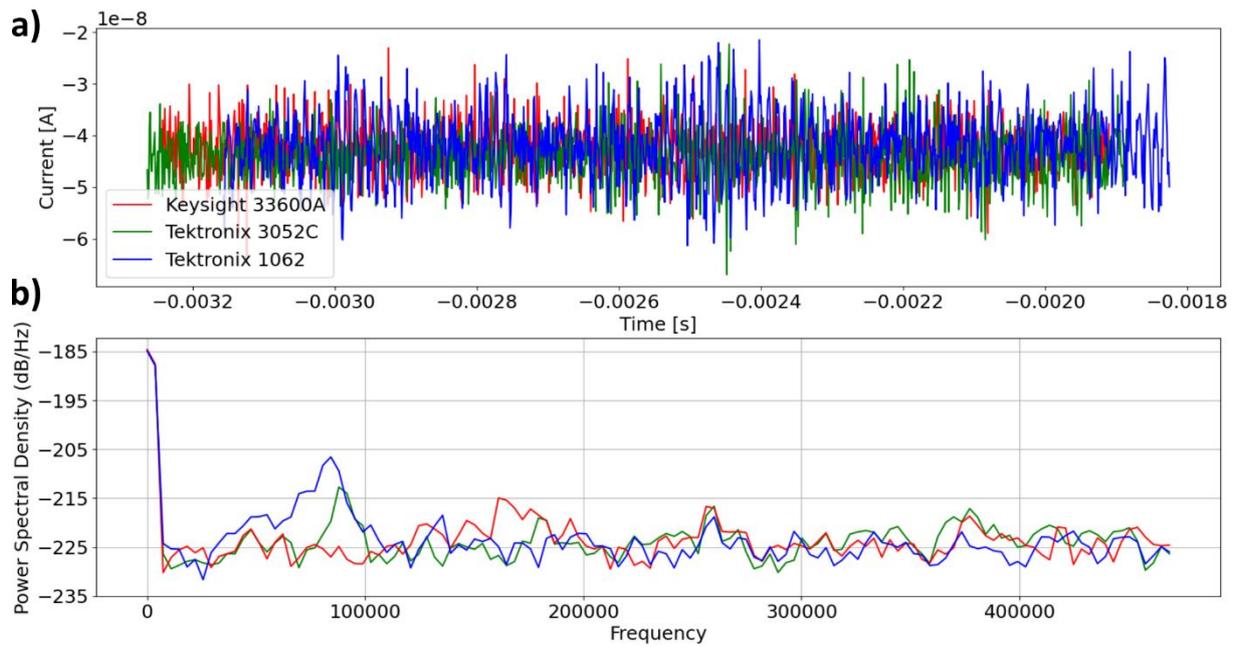

**Figure S3.3:** Noise measurements of the tested AFGs showing **a)** the measured current in the time domain and **b)** the corresponding power spectral density.

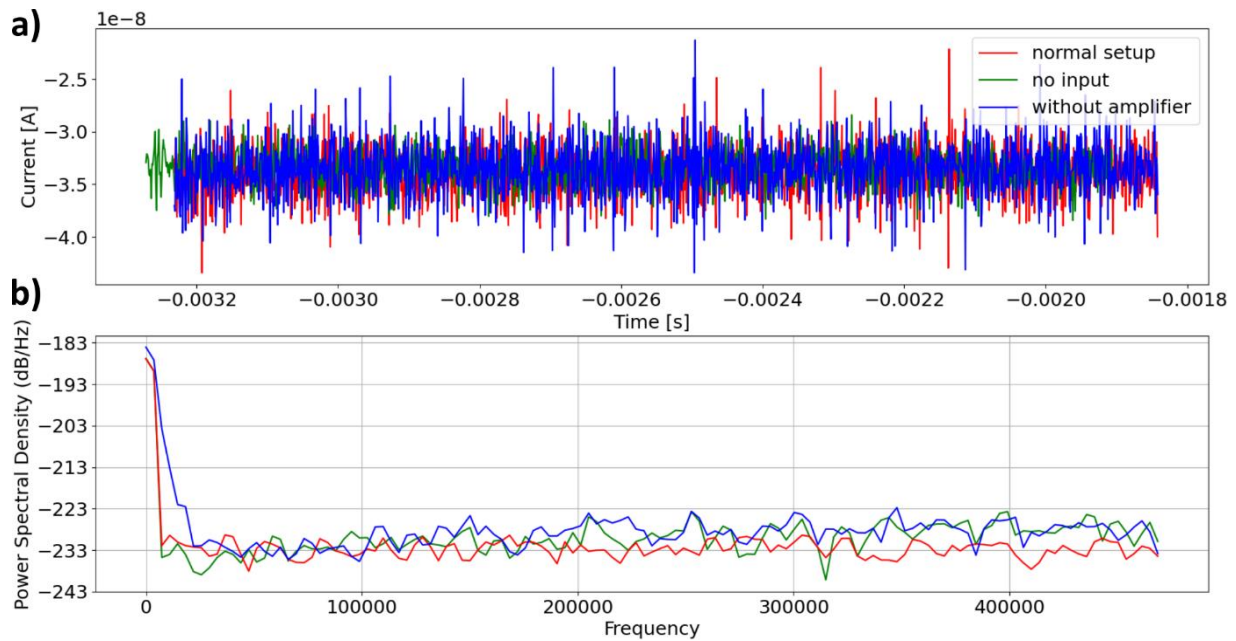

**Figure S3.4:** Noise measurements of the setup with Keysight 33600A Arbitrary Function Generator (AFG) showing **a)** the measured current in the time domain and **b)** the corresponding power spectral density. The noise of the entire setup without a device under test (red line), no input signal (green line, represents the noise floor) and setup without amplifier (blue line) are shown for comparison.

As mentioned in the main text, different rise times $\left((100, 400 \text{ and } 800 \text{ μs})^{+50}_{-0} \text{ μs}\right)$ were investigated. However, it was discovered that the rise times slightly varied within the measurement series. E.g. the 100 μs rise time measurement series encompasses rise times between 100 and 150 μs. The same variability in rise times is present for both 400 and 800 μs. This variation in rise times is attributed to the feedback mechanism of the voltage amplifier reacting to the change in the capacitance of the ferroelectric samples with applied electric field.

## S4. Further experimental data

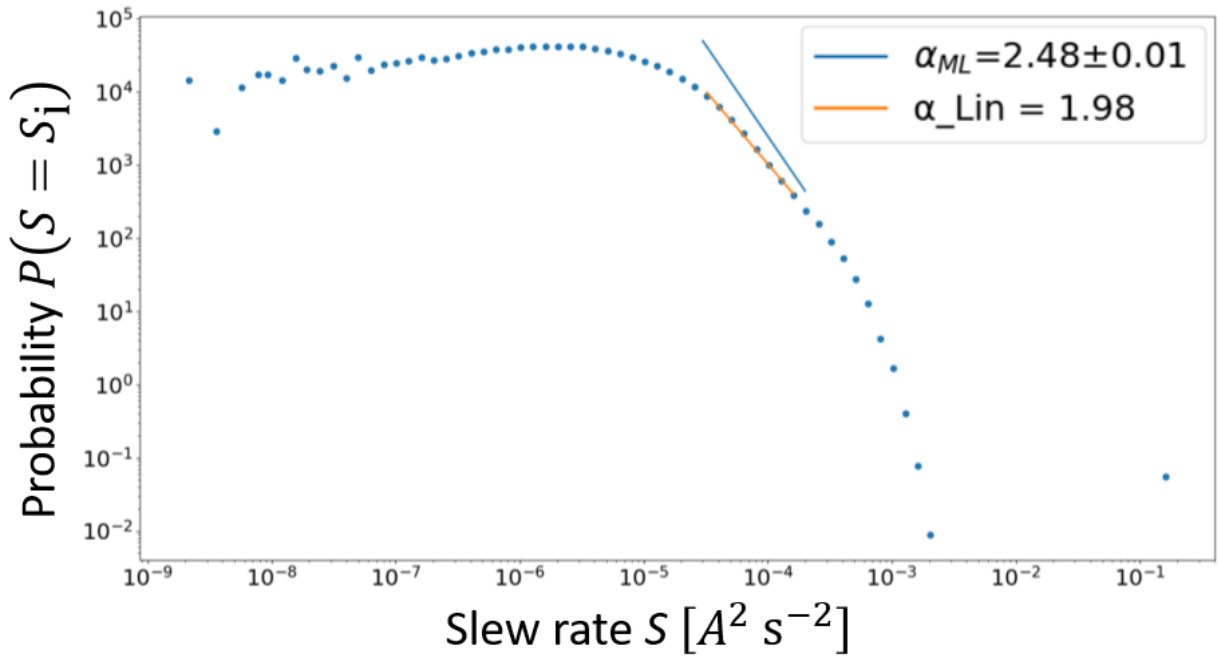

**Figure S4.1:** Measured histogram of the noise spectrum measured for a dielectric reference capacitor at $U = 9.9$ V. The reference capacitor was put into the setup instead of a ferroelectric sample. No power-law behavior is observed over an extended range of slew rates. Instead, a plateau at lower slew rates is apparent, followed by a steep decline appearing as an exponential drop at a slew rate of $\sim 10^{-4} - 10^{-3}$ A$^2$s$^{-2}$. Furthermore, a peak at $\sim 0.4$ A$^2$s$^{-2}$ caused by the displacement current of the capacitor is visible. This reference measurement was used to rule out the possibility that only random noise was measured and was treated as a lower estimate for the noise floor.

**Figure S4.2-4** show the measured histograms of P(VDF:TrFE) noise measurements for the three different rise times $\left((100, 400 \text{ and } 800 \text{ μs})^{+50}_{-0} \text{ μs}\right)$ and different applied electric fields including the fits of the power-law exponents. Some histograms exhibit multiple power-laws and the corresponding fits were done where applicable. Some measurements deviate from the expected behavior which would be a plateau or slow decrease in the probability for lower slew rates, followed by a roll-off, containing the power-law behavior in its beginning part, followed by a steeper exponential drop at the largest event sizes.[11] For example, 19.8 V for 100 μs shows a second bump and 27.2 V for 800 μs initiates the power-law fit within the transitional region. In general, voltages corresponding to electric fields above the coercive field show a more pronounced transition region from the noise plateau (due to thermal fluctuations) at small event sizes to power-law like behavior. This is attributed to the power-law dominating the cut-off region, hence resulting in a direct transition from the plateau to the power-law.

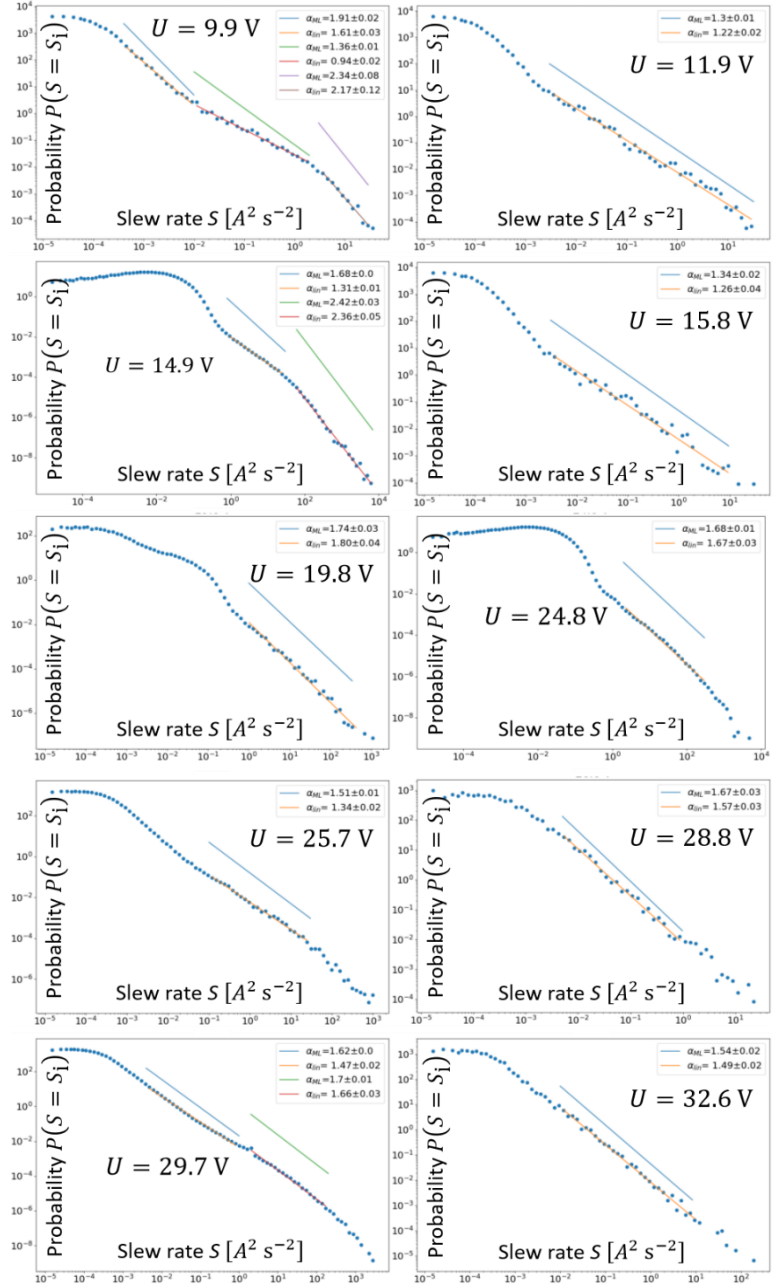

**Figure S4.2:** Measured histograms of the 100 µs P(VDF:TrFE) measurements at different applied electric fields with their extracted power-law exponents. Blue and orange lines are the fitted power-laws using the ML and LS (depicted as 'lin' here) fitting methods. The appearance of multiple sets of fits is due to the distribution seemingly showing multiple regions which may be identified as linear on the double-logarithmic scale. The final fit was chosen based on a combination of manual decision and minimizing the fit error.

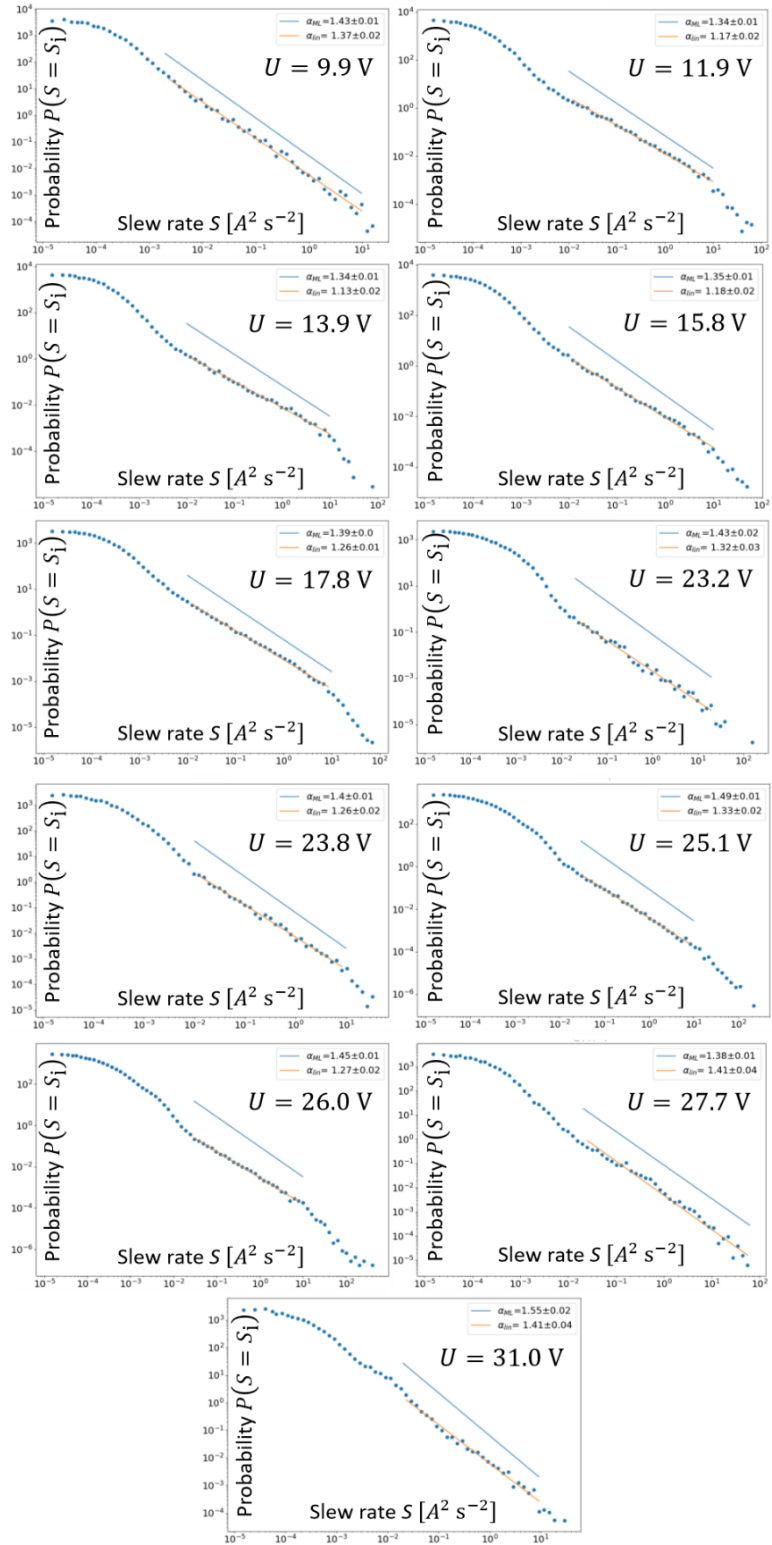

**Figure S4.3:** Measured histograms of the 400 μs P(VDF:TrFE) measurements at different applied electric fields with their extracted power-law exponents. Blue and orange lines are the fitted power-laws using the ML and LS (depicted as 'lin' here) fitting methods.

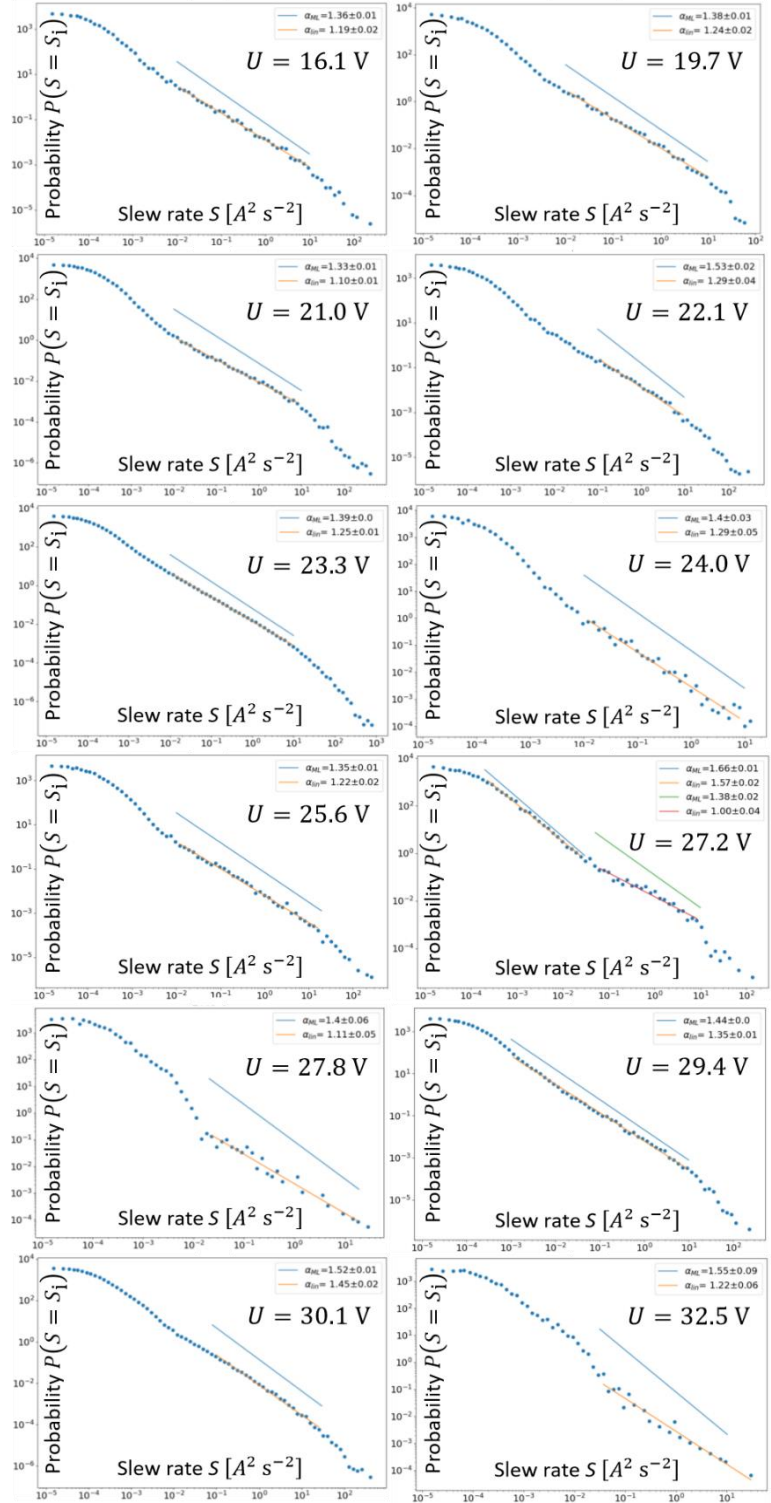

**Figure S4.4:** Measured histograms of the 800 µs P(VDF:TrFE) measurements at different applied electric fields with their extracted power-law exponents. Blue and orange lines are the fitted power-laws using the ML and LS (depicted as 'lin' here) fitting methods.

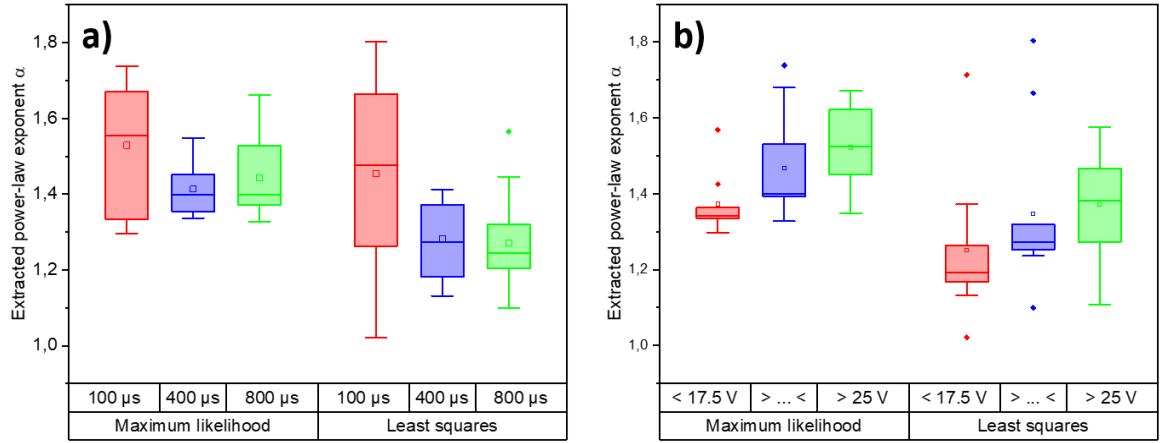

**Figure S4.5:** Box plot of the extracted critical power-law exponents $\alpha$ for **a)** the three different rise times and **b)** three different intervals of applied voltage for both fitting methods including all exponent values obtained from the histograms. The values obtained from ML fitting are slightly larger than from LS fitting for all three rise times, although here the difference between the values for 100 μs is minimal. The trend of increasing exponents with applied voltage is clearly visible.

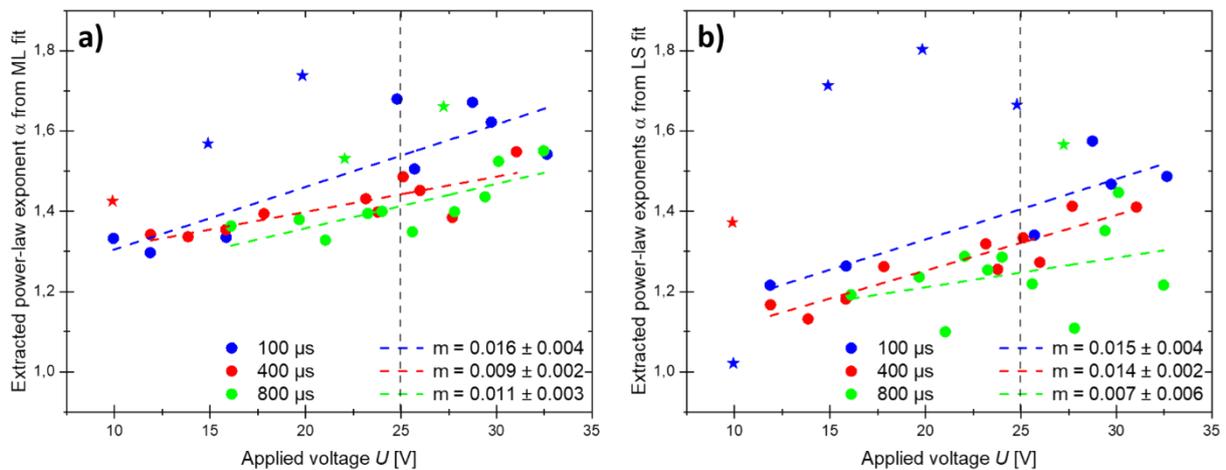

**Figure S4.6:** Power-law exponents $\alpha$ as a function of applied voltage $U$ for three different rise times (100 μs (blue), 400 μs (red) and 800 μs (green)) extracted from **a)** ML and **b)** LS fits of measured histograms. The extracted power-law exponents were fitted by a linear function to illustrate their trends, as indicated by the dashed lines. The corresponding slopes $m$ are included in the plots. Some data points were deemed inconclusive based on the difficulty to obtain unique fits to the histograms and are marked as stars and were excluded from the linear fit. The coercive field of P(VDF:TrFE) is around 25 V as marked by the vertical dashed black line.

**Supplementary references**

[1] J. F. Legrand, *Ferroelectrics* **1989**, *91*, 303.
[2] T. Furukawa, *Phase Transitions* **1989**, *18*, 143.
[3] A. V. Gorbunov, M. Garcia Iglesias, J. Guilleme, T. D. Cornelissen, W. S. C. Roelofs, T. Torres, D. González-Rodríguez, E. W. Meijer, M. Kemerink, *Sci. Adv.* **2017**, *3*, e1701017.
[4] M. Fukunaga, Y. Noda, *J. Phys. Soc. Jpn.* **2008**, *77*, 064706.
[5] J. F. Scott, *J. Phys.: Condens. Matter* **2007**, *20*, 021001.
[6] A. Clauset, C. R. Shalizi, M. E. J. Newman, *SIAM Rev.* **2009**, *51*, 661.
[7] J. Alstott, E. Bullmore, D. Plenz, *PLOS ONE* **2014**, *9*, e85777.
[8] K. Levenberg, *Quart. Appl. Math.* **1944**, *2*, 164.
[9] D. W. Marquardt, *Journal of the Society for Industrial and Applied Mathematics* **1963**, *11*, 431.
[10] M. Poppe, *Prüfungstrainer Elektrotechnik*, Springer Berlin Heidelberg, Berlin, Heidelberg, **2018**.
[11] J. P. Sethna, K. A. Dahmen, O. Perkovic, **2005**, DOI 10.48550/arXiv.cond-mat/0406320.